\documentclass[conference]{IEEEtran}
\IEEEoverridecommandlockouts
\usepackage{cite}

\usepackage[dvipsnames]{xcolor}
\usepackage{amsmath,amssymb,amsfonts}
\usepackage{algorithmic}
\usepackage{graphicx}
\usepackage{textcomp}
\usepackage{xcolor}
\usepackage[]{algorithm2e}
\usepackage{caption}
\def\BibTeX{{\rm B\kern-.05em{\sc i\kern-.025em b}\kern-.08em
    T\kern-.1667em\lower.7ex\hbox{E}\kern-.125emX}}
\begin{document}

\title{Towards a  Real-Time Distributed Feedback System for the Transportation Assistance of PwD}

\DeclareRobustCommand*{\IEEEauthorrefmark}[1]{%
  \raisebox{0pt}[0pt][0pt]{\textsuperscript{\footnotesize #1}}%
}

\author{\IEEEauthorblockN{Iosif~Polenakis, Vasileios~Vouronikos, Maria~Chroni, and Stavros~D.~Nikolopoulos\IEEEauthorrefmark{*}}
\IEEEauthorblockA{\it Department of Computer Science and Engineering \\
 University of Ioannina\\
Ioannina, Greece\\\\
Corresponding Author: \IEEEauthorrefmark{}stavros@cs.uoi.gr}}

\maketitle

\begin{abstract}
In this work we propose the design principles of an integrated distributed system for the augment of the transportation for people with disabilities inside the road network of a city area utilizing the IT technologies. We propose the basis of our system upon the utilization of a distributed sensor network that will be incorporated by a real-time integrated feedback system. The main components of the proposed architecture include the Inaccessible City Point System, the Live Data Analysis and Response System, and the Obstruction Detection and Prevention System. The incorporation of these subsystems will provide real-time feedback assisting the transportation of individuals with mobility problems informing them on real-time about blocked ramps across the path defined to their destination, being also responsible for the information of the authorities about incidents regarding the collision of accessibility in place where the sensors detect an inaccessible point. The proposed design allows the addition of further extensions regarding the assistance of individuals with mobility problems providing a basis for its further implementation and improvement. In this work we provide the fundamental parts regarding the interconnection of the proposed architecture's components as also its potential deployment regarding the proposed architecture and its application in the area of a city. 
\end{abstract}

\begin{IEEEkeywords}
Smart Cities, Sensor Networks, Disabilities, Distributed Systems 
\end{IEEEkeywords}

\section{Introduction}
In the context of everyday living and moving in the urban web, unfortunately, it is not uncommon for us to see many of our fellow human beings deprived of the right to move at a time when some drivers sometimes consciously and sometimes do not choose to park their vehicle in front of a a transition point that serves people with mobility difficulties. As a daily phenomenon, the implementation of the proposed system aims to facilitate the daily movement of people with mobility difficulties in relation to the violations that occur during the violation of the Highway Code and consist of the exclusion of crossings intended for use by persons of this vulnerable social group. 

In this work, an integrated distributed system is presented, an integrated system of safe movement in the city in order to improve the daily life of people with mobility difficulties. The structure of the proposed system for the safe movement in the city of people with mobility difficulties, consists in the development of three axes, in terms of informing users (people with mobility difficulties) about points that have been excluded, the analysis of relevant data that will arise from the previous process, and the further development of a method of preventing such phenomena.

\subsection{Transportation Assistance of PwD}

Transport services for PwD is a factor that cuts across the main concern of this journal in that helps to increase participation and independence. Accessible transportation is crucial for members of their community who acquire disabilities, to address vital aspects such as: attending educational facilities, access and exercise of employment opportunities, as well as access to social, and recreation. These services include public transport solutions for individuals with disabilities, other methods of transport that incorporate disability and services for funding transport expenses. Transportation planning should extend to accounts that consider access to modes of transport and associated facilities, automobiles, frameworks, information technology interface designs, and other parameters. Subsequently, modern transport systems have enhanced the accessibility of PwD through improved transport systems around the globe.

In the present days, most cities use low-floor buses, elevators or ramps in subway stations, and the tactile pavings to help the visually impaired. Also, other forms of transport, and particularly the ride-sharing services, have slowly allowed the use of wheelchair-accessible vehicles (WAVs) in the market hence improving the flexibility among PwD. Nevertheless, major issues persist as to how one could secure proper transportation accessibility for everyone out there. Accessibility to timely, appropriate, and affordable transport is particularly a problem in rural settings, where people with disabilities are required to make very few trips the costs of adaptations may also be higher. In addition, it has been noted that the accessibility and quality of transport services that individuals with various disabilities may use can also differ significantly depending on the region. In order to address these issues, development of better transport infrastructure, better qualified staff in transport sector and proper funding to transport helper services and facilities are needed. Human society needs to dedicate more of its resources to focus on how it can enhance the mobility requirements of the PwD to ensure a barrier-free lifestyle for such individuals.

\subsection{Real-time Systems in Assistance of PwD}
Advanced real time systems to help the disabled change the direction of accessible transport services to persons with disability. These systems incorporate smart technologies such as Global Positioning System (GPS), big data, and the concept of things (IoT) for designing a dynamic transport solution. Through such technologies, the system provide for efficient use of accessible vehicles when they are dispatched and directed so that they may meet the target serve the intended clients with less waiting time and increased reliability. For instance, the system can monitor traffic, receive latest updates to the database, and determine the most optimal routes taking into consideration traffic density and compliance to timings for pick-ups and drop-offs of PwD.

Considering the main advantages of a Real-Time Distributed System one must mention that they can deliver individual and instant aid and assistance to the users. This is because choosing an automobile to use, ordering the automobile, indicating the need for additional facilities like wheel chair or help in boarding, and even monitoring the automobile as it looks for the customer and the time it will take to arrive at the required location, can all be done through mobile application. This level of interactivity and customization does not only enhances the use of the apps but also enhances the independence of PwD in selection of mobility. Moreover, the system can inform drivers and support staffs on the specific requirement prior to check-in so that accommodation is arranged in time and there would be no obstacles to accommodate the need of the patient.

In addition, the application of a Real-Time Distributed System enhances the availability of the transport system by diversifying the system or means of transport. Thus, more extensive information regarding service usage and attitude to accessibility can be shared, and transportation authorities can plan the additional distribution of resources and network redesign. This is a fluid data-oriented approach that can be used in enhancing an organization's performance through streamlining service provision to PwD. Besides, due to the system scalability, it can easily expand its reach to cover a bigger area of the country and add new technologies in the course of its operation. Since this generates methods that premier the mobility of PwD, society is inching closer to the concept of a mobility impaired community that could gain access to transport solutions as other people do.

\subsection{Related Work}

Tavares et al. in \cite{TaBaCaCoYaRe}, propose Hefestos, a concept for ubiquitous accessibility, was created to support PwD without regard to the nature of their disabilities. The users' mobile devices can be used to find accessible resources in a particular area. The user moves while being informed of the resources that are nearby and is able to reach them thanks to this information. Kbar et al. in \cite{KbAlElBhAlEn} utilize Ambient Intelligence (AmI) elements that are based on assistive technology, to support PwD at work by SHWPWD, where the propose approach allows people with disabilities to use products with little effort and a high degree of comfort, and it also helps them to carry out various tasks in the workplace more successfully. Aly in \cite{Aly} based on smart-phones and Wireless Sensor Networks (WSN) proposes MNDWSN aiming in the in indoor assistance of blind, deaf, and disabled individuals. Fernandes et al. in \cite{FeFiCoBa}propose the navigation system LBSBlind [19] to help visually impaired persons and provide them security while they are in motion. Using the Global Positioning System and Radio-Frequency Identification (RFID) as its primary navigational systems, LBSBlind attends to both indoor and outdoor situations (GPS). 

Wei et al. \cite{Wei}, propose a smart set containing a smart cane and headset that guides people with visual difficulties. Apart from the smart cane and headset, the authors developed also a smart app that integrates the above smart set.The blind individual uses the guide app to voice-input their destination following that, the app uses the GPS map system to compute the best route for them. The app then broadcasts the direction in real-time through the bone conduction headset for guiding. Huang et al. in \cite{Huang}, propose a sytem namely SAPSS for the visually-impaired. SAPSS is a combination of hardware, software, and user-side mobile application, and consists of two parts, the hardware part which is the communication box attached to existing pedestrian traffic lights and the mobile app part that provides accurate geo-location positioning of the users and real-time feeding of pedestrian signals to their smart-phones. The mobile app is connects through Bluetooth with the traffic light communication box. Wheeler et al. in \cite{Wheeler}, propose an open source data model namely ADE-AP that sits above the CityGML data model and provides alternative pathways for people with disabilities. ADE-AP is a model that bridges the gap between non personalized pathways for people with disabilities and the non open source platform from where someone can access those data. 

Patwary et al. in \cite{Patwary}, propose an intelligent sign language recognition system for dumb people, build in an wearable glove with flex sensors.The system consists of  an Arduino Uno, an LCD screen, a GS M 800L module, and lithium-ion rechargeable battery. The Arduino Uno microcontroller senses this voltage when the sensor turns the detected execution load into a voltage. Finally, the detected signs are displayed on the LCD to display and pronounced verbally by the speaker. Tegeltija et al. in \cite{Tegeltija}, propose a solution that relies on cars who park in spots designated for people with impairments being recognized. The described system ensures user validation by detecting when a parking space is occupied by a parking sensor and by validating the user through a mobile app after scanning a QR code. If the user is not permitted to park, the system notifies the necessary service. 

\subsection{Motivation and Contribution}
An observation that is seen everyday in big urban regions, is the obstruction of PwD transition points and the deprivation of the people with mobility difficulties to move freely. The obstruction of PwD points either being consciously made by drivers or not, can lead to some serious situations such as people with mobility difficulties may choose alternative routes that are unsafe. Apart from the danger of choosing unsafe alternative routes, the psychological frustration can be a problem to as the freedom of moving freely is deprived.

In this work we propose an integrated distributed sensor network that will detect and prevent the obstruction of PwD transition points. The propose system will be a real-time assisting system that will help people with mobility problems to choose safe routes through the city as it will receive in real-time the obstructed points and recalculate an alternative route. Moreover the proposed architecture incorporates different technologies both from the field of sensors and the field of software, making it an IoT navigating system for people with mobility difficulties.%

\section{Theoretical Background}
In this section we discuss the theoretical background regarding the basis of our model concerning the sensor networks, the client-server applications and the integrated cyber-physical systems.

\subsection{Sensor Networks}
Sensor networks performing the tasks of  processing, analyzing, and disseminating data enable access to information at any time and from any location \cite{TuMa}. Our proposed system is composed by sensors that are connected through wireless links in order to transfer data regarding the existence, or not, of obstacles in front of them, i.e., ramps for lowering disabled people from a pedestrian sidewalk are blocked by illegally parked cars. These data are then transferred into a sever that registers this information while interaction with a client application informs the used whether the ramp is blocked by obstacle. These environment constitutes a cyber-physical system where sensors and high-end computers interact by performing particular procedure in real-time in order to inform the users about the reachability of particular points in the city that are part of their paths followed through a transition among different points in the city.

\subsection{Integrated CPS for Detecting Blocked Wheelchair Ramps}
The proposed approach utilizes the development of an integrated cyber-physical system for the detection of wheelchair ramps that have been blocked by illegally parked vehicles. The design and the development of the proposed system utilized the Arduino Uno micro-controller board, which is based on the Microchip ATmega328P, and the utilization of several sensor modules attached to it in order to derive an integrated system for the detection of blocked ramps. In particular there are utilized the HC-SR04 Ultrasonic Range Sensor, the Ks0232 Red LED Module, the KS0018 Active Digital Buzzer Module, and the ESP8266 WIFI Module. 

\begin{figure}[t!]
\centering
\includegraphics[scale=0.46]{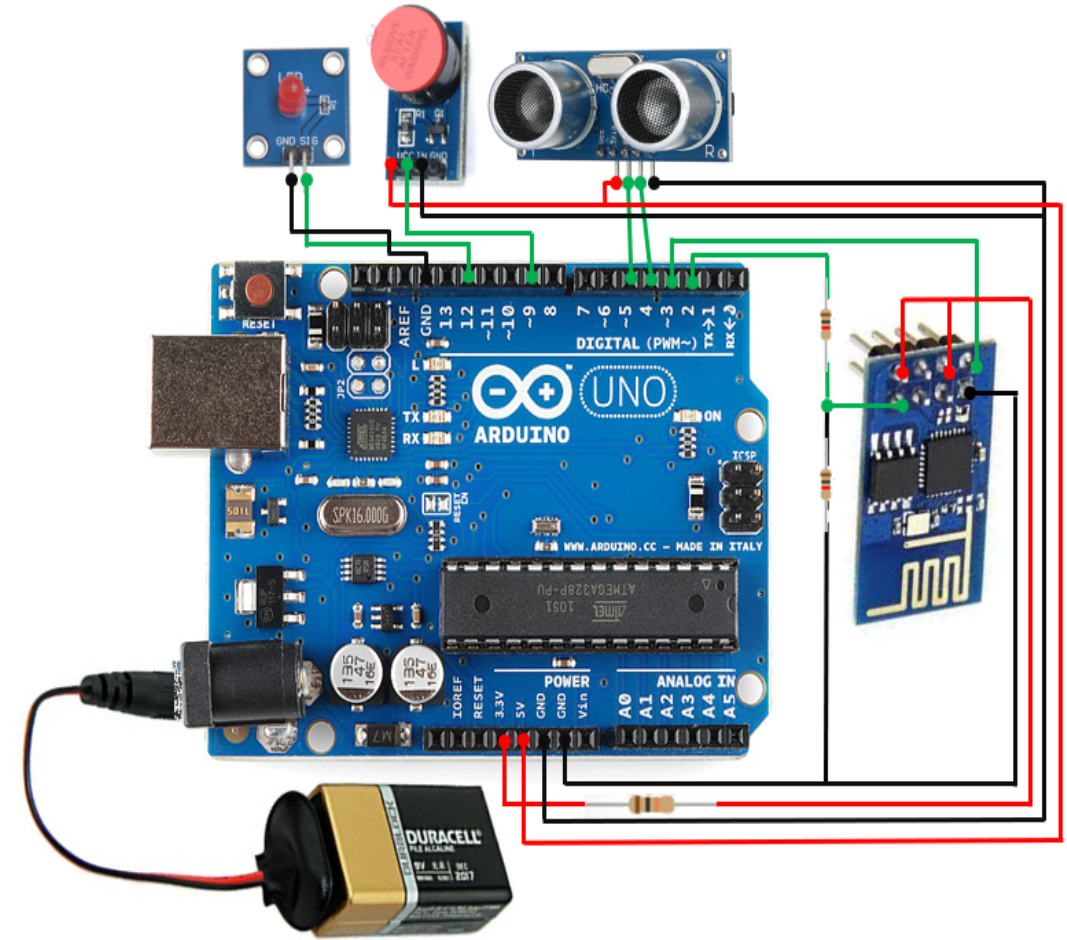}
\caption{Wheelchair ramp sensor prototype.} \label{arch1}
\end{figure}

In Figure~\ref{arch1}, it is illustrated the architecture followed for the integration of the modules required for the procedure of detecting whether a wheelchair ramps has been blocked, into a unified system. To this point we should note that, as we will discuss later, there are two modules, i.e., the Red LED and the  Buzzer modules, that do not operate directly on the task of detecting whether a wheelchair ramp has been blocked by a parked vehicle, but instead they are activated producing a characteristic audiovisual effect when an obstacle has been stable in the detection area of the ultrasonic range sensor more than a specific time interval, e.g., in our case, $3$ minutes.

\subsection{Real-time Distributed Feedback}
The proposed architecture relies upon the distributed management of the collected information by both the users of the system themselves as also by the sensors. In particular, based on a client-server architecture there are distinguished two types of clients, namely the standard users of the system, i.e., PwD that make use of the mobile version of an application that provides information incorporating the underlying feedback system, as also the standard sensors that are placed in front of the wheelchair ramps. In the case study considered into this work, we take into account the first scenario where the users (i.e., PwD) utilize a mobile application in order to either mark inaccessible points (e.g., points in the sidewalks where obstacles are blocking a pass through) or to be informed about the wheelchair ramps that have been blocked in order to prepare and modify their routes accordingly. On the other hand, the second scenario considers the utilization of the integrated CPS that deploys the procedure of blocked ramp detection and prevention by informing both the system to derive consequently this information to the users, as also to the individual that blocked the ramp about the event.
\begin{figure}[t!]
\centering
\includegraphics[scale=0.54]{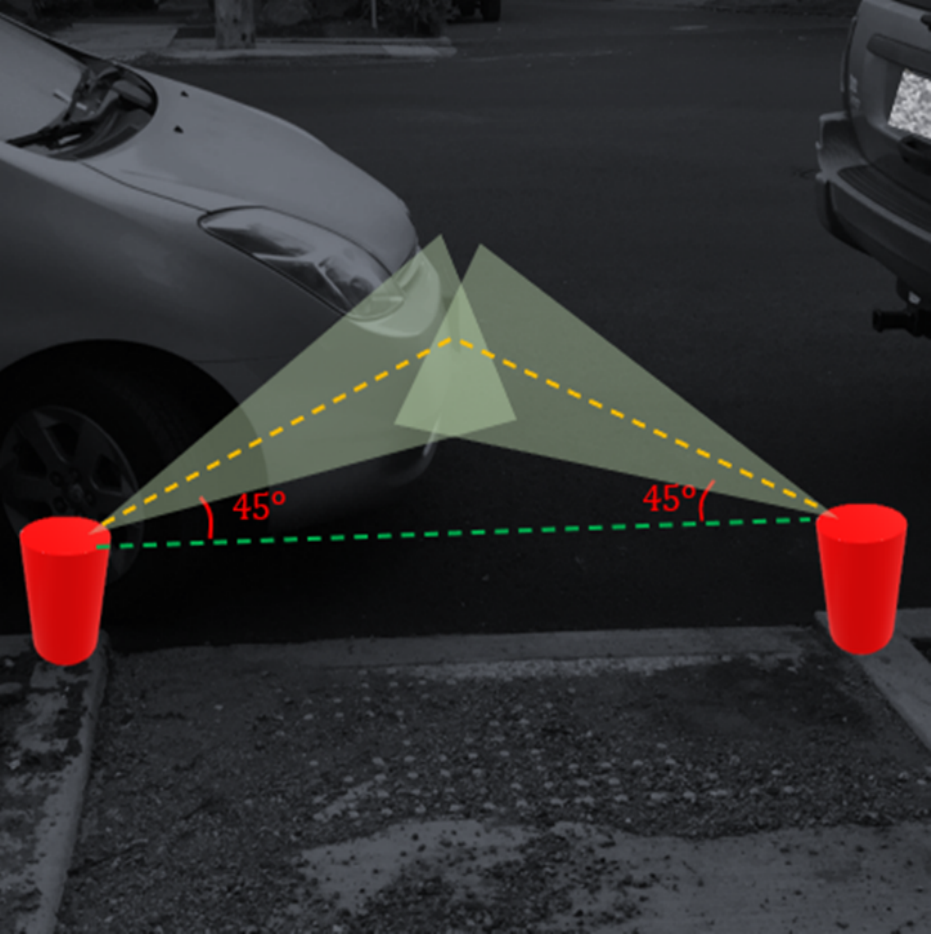}
\caption{Example of \textit{Semi-blocked} wheelchair ramp.} \label{arch}
\end{figure}

\begin{figure}[t!]
\centering
\includegraphics[scale=0.54]{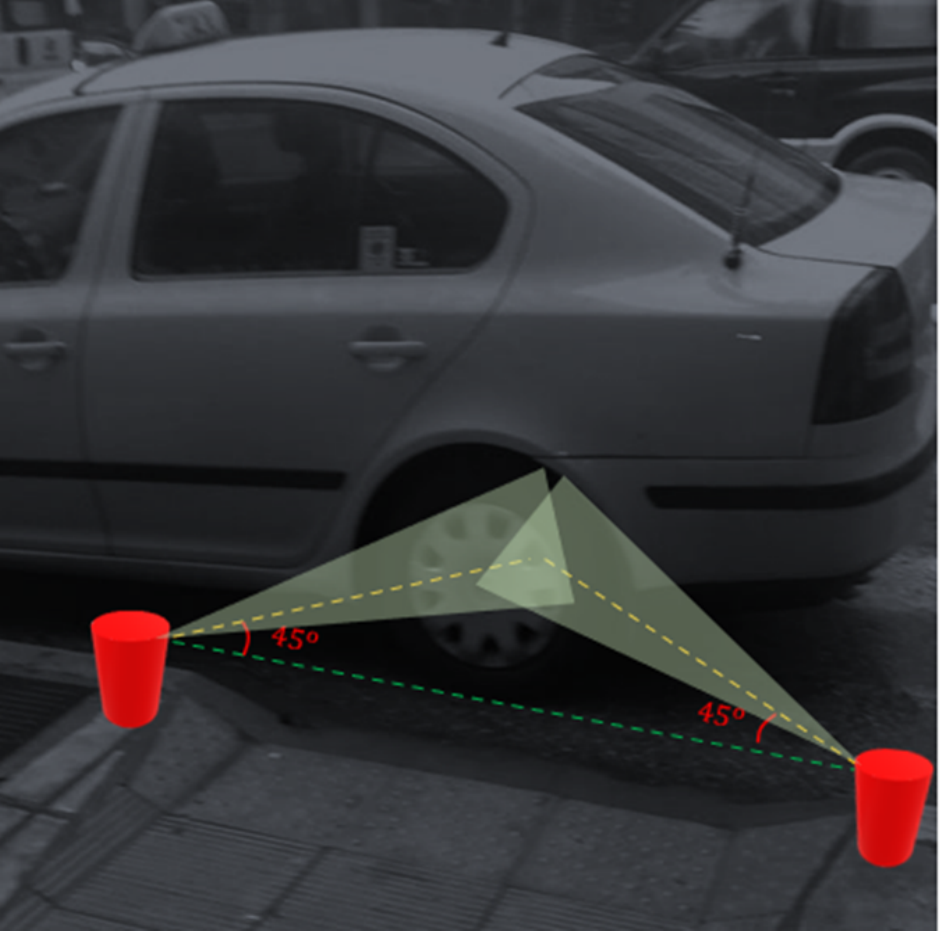}
\caption{Example of \textit{Full-blocked} wheelchair ramp.} \label{arch}
\end{figure} 

\begin{figure*}[t!]
\centering
    \begin{minipage}[t]{.48\textwidth}
        \centering
        \includegraphics[width=1\textwidth]{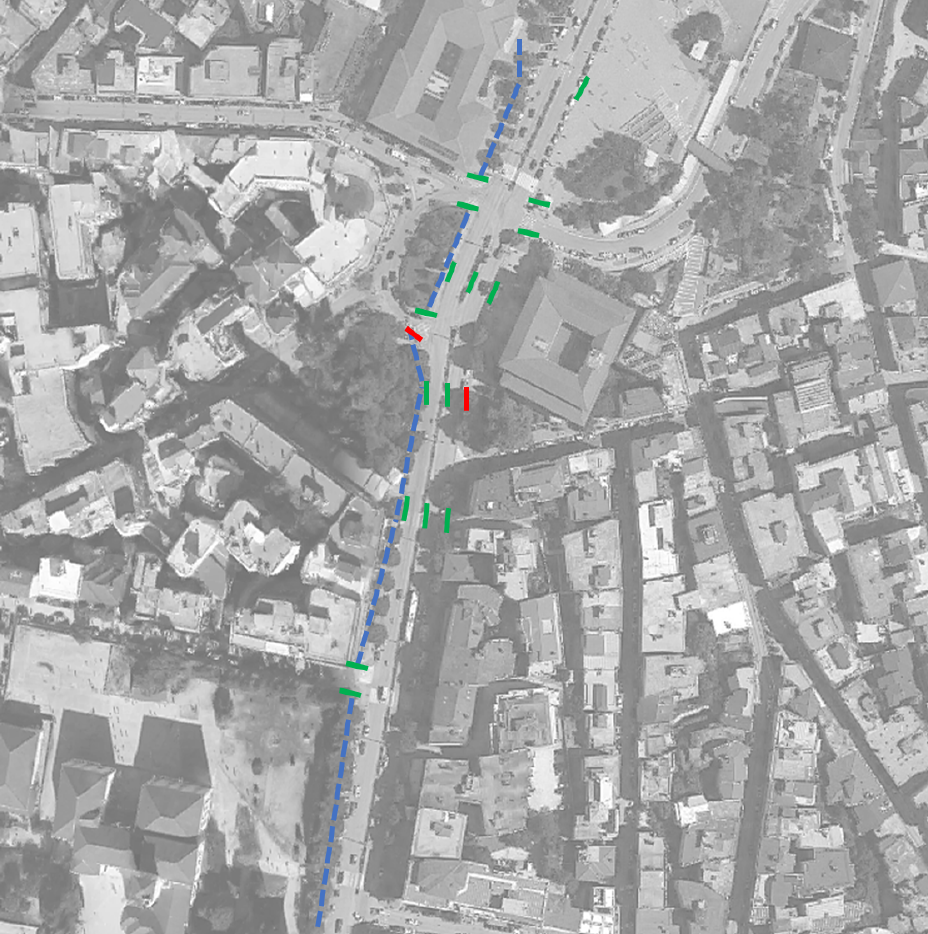} \\
         (a)
    \end{minipage}
    \hfill 
    \begin{minipage}[t]{.48\textwidth}
        \centering
        \includegraphics[width=1\textwidth]{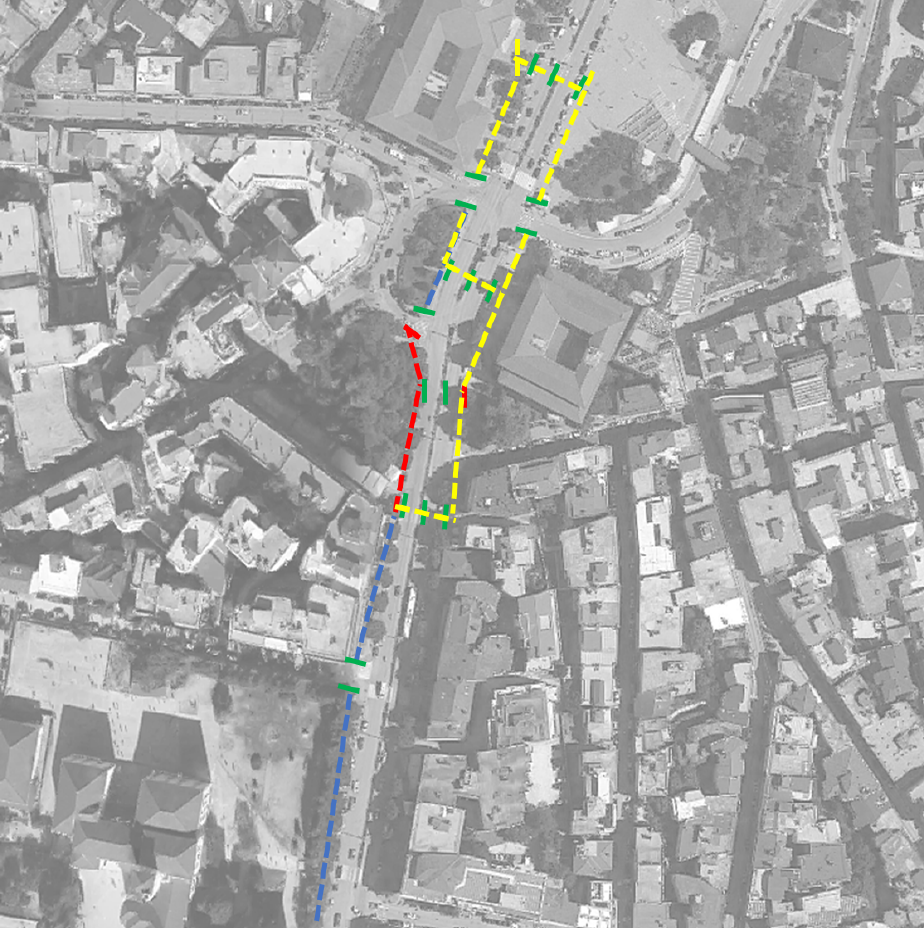}\\
         (b)
    \end{minipage}  
    \label{fig:1-2}
    \caption{Example of original (a) and alternative (b) route in presence of \textit{Semi-blocked} or \textit{Full-blocked} wheelchair ramps.}
\end{figure*}

\section{System Architecture and Deployment}

\subsection{Architecture}
The architecture proposed in this work is composed of three subsystems that make up an ecosystem based on the client-server model. Namely, the components that make up the propose architecture are: \\

\begin{enumerate}
\item subsystem for marking inaccessible points, \\
\item data flow analysis subsystem, and \\
\item obstacle prevention subsystem. \\
\end{enumerate}

On the part of the beneficiaries, the users (clients) will utilize the application that will run on their mobile devices (mobile app) by receiving and sending information to the central system, thus creating a distributed model of pumping and providing information regarding the situation at the points. access for people with mobility difficulties. On the other hand, in the server part and regarding the back-end of the application, the integrated system will pump and provide information to the individual subsystems, and in addition will take over the management of the information flow both through the users and through the layout sensors, where they will be used by each subsystem.

Specifically, through the distributed architecture of the proposed system we can observe the data flow between the entities and the interaction between them, based on the client-server model where both the flow of information from the users of the application to the system and the system to users thus providing a distributed real-time two-way flow feedback framework. In this context, it becomes clear that the application, through the integrated system, will be able to supply its users in real time for places in the city that are inaccessible, whether they have been recorded through the sensor network or by others. users, thus delimiting a passive model to help the movement of people with mobility difficulties, while at the same time, at an active level, to immediately notify the competent authorities to ensure that these points are re-accessible.

\subsection{Deployment}
Next we discuss the component deployed in our proposed system, namely the MICPS, the LDARS, and the ODPS.

\vspace{0.2 in}
\noindent\textbf{1. Mark Inaccessible City Point System - \textit{MICPS}}

\noindent Specifically, in the first axis we have the development of the network and mobile version of the \textit{MICPS} sub-system application that aims at its use by people with mobility difficulties:\\

\begin{itemize}
\item marking on a digital interactive map of points of the city (which are intended for the passage of persons with mobility difficulties) and in which their passage is temporarily hindered (eg from a parked vehicle) as soon as the persons are in front of such a point, and \\
\item informing in real time the rest of the members of this social group who intended to pass through that point, in order to avoid their unnecessary movement.\\
\end{itemize}

The aim of this subsystem is to create an interactive map of the city with access points for people with mobility difficulties in which traffic is not possible. This subsystem, as well as the corresponding mobile application will protect people with mobility difficulties by giving them the opportunity to actively inform on the interactive map for inaccessible points, while on the other hand, at the level of passive information, they will be provided as soon as possible. the choice to adjust their route avoiding painful transitions and unnecessary inconvenience until they reach a point (e.g., ramp) finally realizing that at that point there is a parked vehicle that impedes their passage.

\vspace{0.2 in}
\noindent\textbf{2. Live Data Analysis and Response System - \textit{LDARS}}

\noindent In a second level, the implementation of the LDARS subsystem will utilize the information that will be integrated in real time in the interactive digital map by the application of MICPS, in order to analyze data regarding the "inaccessible" points of the city and consequently to start the immediate feedback of the system to the community. The main objectives of the \textit{LDARS} subsystem will be: \\
\begin{itemize}
\item the statistical study of the frequencies of occurrence and geographical characteristics of the points that appear with increased frequency as ``inaccessible", and \\
\item informing the competent authorities in real time so that they can rush to the specific point as soon as possible \\
\end{itemize}

In the long run, such statistical data will give the opportunity to the region as well as to the competent bodies to properly address the implementation of the foreseen legal and institutional frameworks in order to holistically solve the problem of travel disability for people with mobility difficulties.
%
%
%

\vspace{0.2 in}
\noindent\textbf{3. Obstacle Detection and Prevention System - \textit{ODPS}}

\noindent As an extension of the above, and towards the completion of the system, the proposed architecture integrates the \textit{ODPS} subsystem which aims to actively prevent drivers who decide to park their vehicle in front of a special crossing point for people with mobility difficulties. Specifically, this subsystem, through the use of ultrasonic sensors built into micro-controller platforms with the possibility of wireless connection to the Internet can recommend integrated systems which will be integrated in the special access points for people with mobility difficulties, with aims: \\

\begin{itemize}
\item the detection obstacles (e.g., illegally parked vehicles), and \\
\item provide immediate updates of the interactive map of the \textit{MICPS} subsystem. \\
\end{itemize}

Also, it is worth mentioning that the placement of these integrated micro-controller structures can be achieved by utilizing the data collected during the implementation of the \textit{LDARS} subsystem where through it can be selected the placement of the respective structures in parts of the city that have a high frequency of instruction. map being in ``inaccessible" mode.

\begin{figure*}[t!]
\centering
\includegraphics[scale=0.84]{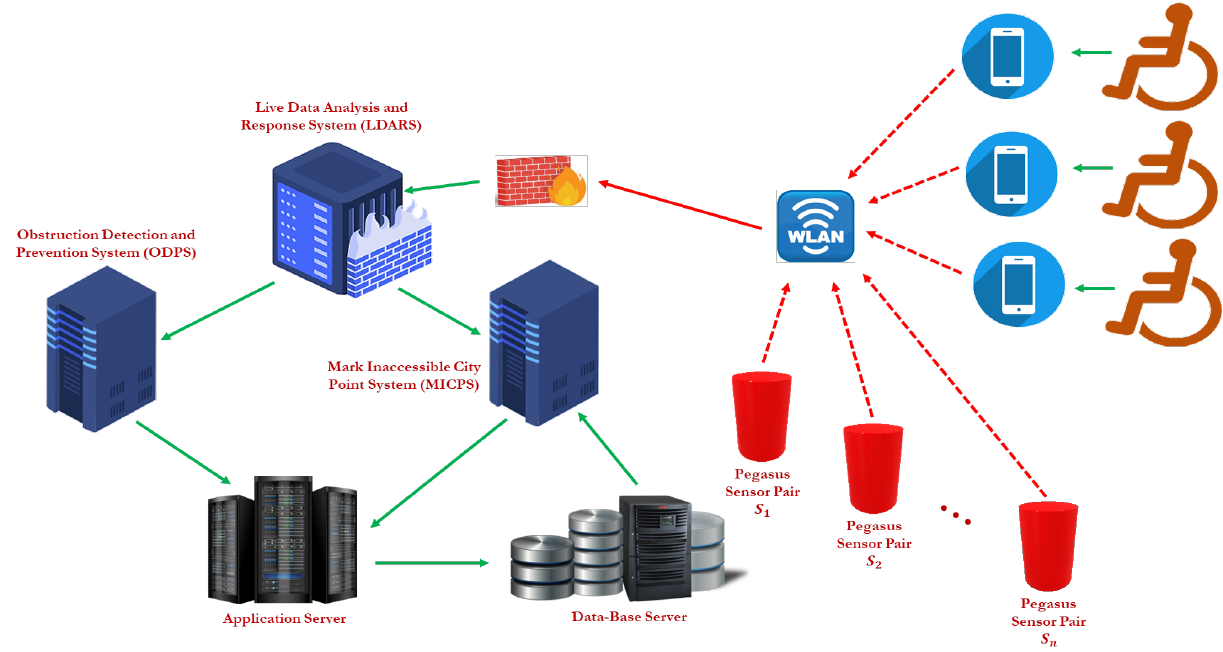}
\caption{System Architecture.} \label{arch}
\end{figure*}

\subsection{System Integration}
Regarding the adaptation of the system to the level of material requirements, it is worth mentioning that the costs required for the implementation of an integrated distributed system can be limited solely to the acquisition of the equipment required to build the integrated sensor systems required to be installed in order to become operational. The process of detecting and preventing the creation of obstacles in access points for people with mobility difficulties that is implemented through the \textit{ODPS} system. 

Respectively, for the other equipment required for the implementation of the integrated distributed system, regarding the network infrastructure and the use of servers, the costs can be reduced by the integration, initially of a pilot application, in the material and technical infrastructure of the region. , with the immediate benefit of utilizing existing resources, and consequently the further savings of installation and maintenance costs that may have resulted from the use of exclusive and new logistical equipment.

\begin{figure}[t!]
\centering
\includegraphics[scale=0.44]{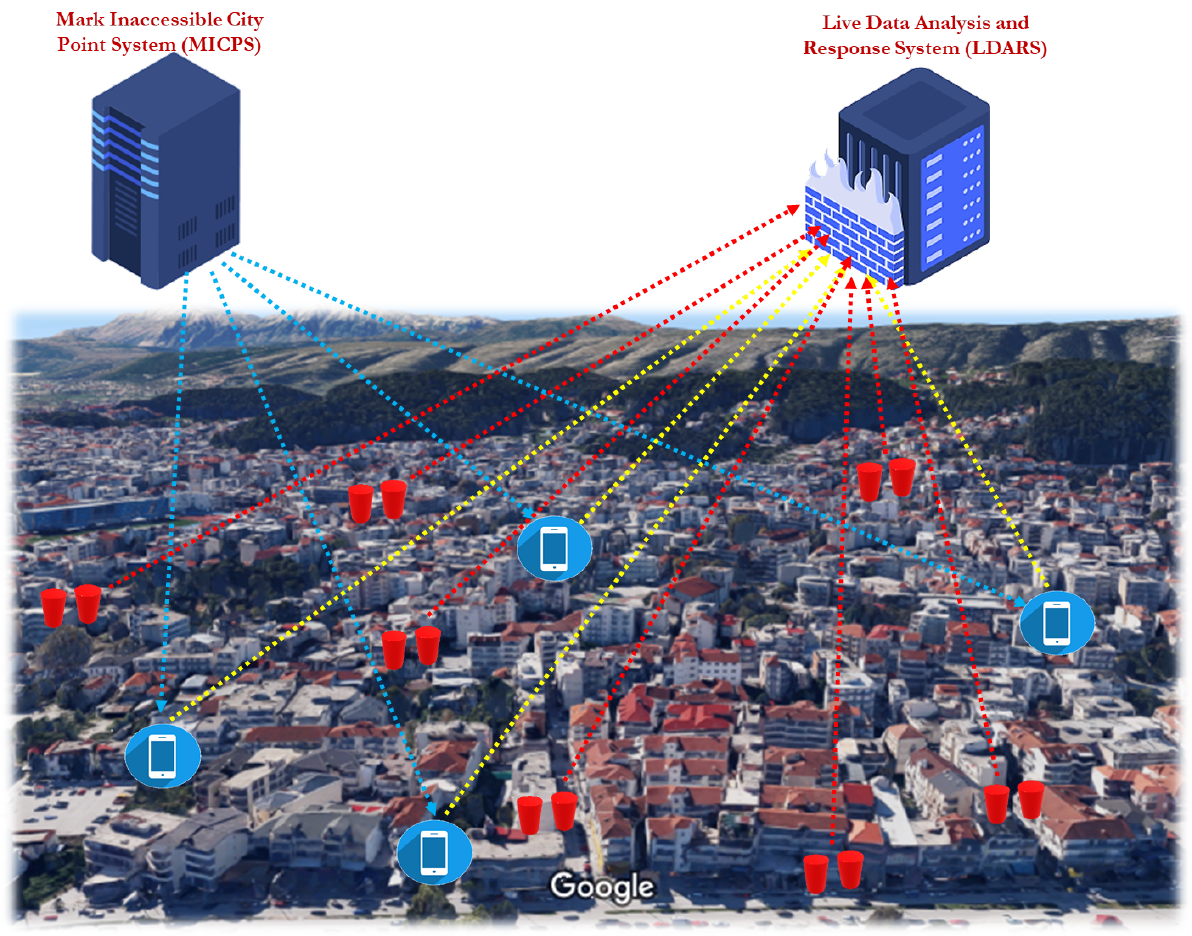}
\caption{System Deployment.} \label{arch}
\end{figure}

\section{Conclusion}
Next, we discus the concluding remarks regarding the benefits the could be exhibited by the application of the proposed distributed system, the integration issues that may arises, as also the potentials that are provided by the deployment of the proposed system.
 
\subsection{Potentials ans Limitations}
The implementation of the proposed real-time system for the safe movement in the city of people with mobility difficulties is expected to provide an immediate perspective in improving their daily lives. This proposal is expected to provide a direct interface between the system and the users of the application, users of the application with other users of the application, but also the users of the application with the state itself. 

In particular, the proposed distributed two-way real-time data flow model will ensure, to a large extent, in addition to providing immediate mobility assistance to people with mobility difficulties, the link between person and state in order to enhance the link between these entities, allowing application of modern technologies in solving sensitive everyday problems.

\subsection{Remarks}
Finally, it is worth mentioning that the prospects and viability that exist from the development of such an application are not limited to the operation of the application itself. In particular, given the modular arrangement of the subsystems that make up the distributed architecture of the proposed system, it is obvious that further subsystems can be added using the same hardware infrastructure as the one used by the proposed system integrating all other components (e.g., traffic congestion control) on the same existing platform. 

This capability ensures to a large extent the viability and scalability of the proposed system by defining a single axis of technological development that will actively connect the local community with the state and the competent regional authorities to address daily transportation problems in the city, not just individuals. with mobility difficulties but also of other citizens.


\end{document}